\documentclass[twocolumn,secnumarabic,amssymb, nobibnotes, aps, pra]{revtex4-1}

\usepackage{framed}
\usepackage{amsfonts}
\usepackage{amsmath}
\usepackage{graphicx}
\usepackage{color}

\usepackage[hidelinks]{hyperref}
\usepackage{xcolor}
\hypersetup{
	colorlinks,
	linkcolor={blue!80},
	citecolor={blue},
	urlcolor={blue!80!black}
}

\begin{document}
\title{Dark quadratic localized states and collapsed snaking in doubly resonant dispersive cavity enhanced second harmonic generation}
\author{P. Parra-Rivas, C. Mas Arab\'i, F. Leo}

\affiliation{ OPERA-photonics, Universit\'e libre de Bruxelles, 50 Avenue F. D. Roosevelt, CP 194/5, B-1050 Bruxelles, Belgium\\
}

\date{\today}


\begin{abstract}
We theoretically investigate the dynamics, bifurcation structure and stability of quadratic dark localized dissipative states arising in cavity enhanced second-harmonic generation. These states, formed through the locking of plane front waves, undergo collapsed homoclinic snaking and may exhibit oscillatory dynamics.
\end{abstract}
\maketitle

\section{Introduction}
Localized dissipative structures (LSs) appear in a large variety of natural domains, such as population dynamics, plasma physics, solid mechanics and nonliner optics \cite{descalzi_localized_2011,akhmediev_dissipative_2005}. Their formation is in general related with the coexistence of different stable extended states within a parameter range of the system, and the locking of front waves connecting them \cite{pomeau_front_1986,coullet_nature_1987}.

In optics, these states may arise in externally driven nonlinear cavities, where light can be trapped and interact continuously with a nonlinear medium. Light localization in dissipative systems is hold by a pairwise equilibrium where nonlinearity compensates dispersion and/or diffraction, while energy dissipation is balanced through external energy driving.   
Light LSs have been shown experimentally in cavities with Kerr type of nonlinearity, as for example in semiconductor microcavities \cite{barland_cavity_2002}, fiber cavities \cite{leo_temporal_2010}, and in dispersive microresonators \cite{herr_temporal_2014}. Here, LSs have been proposed for different technological applications such as information processing or metrololy.

In quadratic cavities, LS formation has been intensely studied, mainly in the context of  optical parametric oscillators (OPOs), both in diffractive \cite{staliunas_localized_1997,trillo_stable_1997,longhi_localized_1997,tlidi_spatiotemporal_1997-1,tlidi_kinetics_1998,staliunas_spatial-localized_1998,staliunas_three-dimensional_1998,oppo_domain_1999,tlidi_space-time_1999-1,berre_striped_1999,skryabin_instabilities_1999,tlidi_three-dimensional_1999,oppo_characterization_2001,rabbiosi_new_2003} and dispersive cavities \cite{parra-rivas_localized_2019,parra-rivas_parametric_2020,nie_quadratic_2020,nie_quadratic_2020-2}.
Cavity enhanced second-harmonic generation (SHG) has received, however, much less attention than OPOs. In diffractive cavities Etrich {\it et al.} shown that LSs could appear in the scalar (i.e., Type I SHG) \cite{etrich_solitary_1997,michaelis_quadratic_2003}, and vectorial (i.e, Type II) configurations  \cite{michaelis_quadratic_2003,longhi_spatial_1998,peschel_symmetry_1998}. Here, LSs form in the transverse plane to the propagation direction, and they are two-dimensional structures. Moreover, in the last case three-dimensional LSs were also theoretically found \cite{tlidi_three-dimensional_1999}.

Just a few years ago, quadratic dispersive cavities gained a lot of attention due to their potential application for frequency comb generation \cite{leo_walk-off-induced_2016,leo_frequency-comb_2016,hansson_singly_2017}. In this context, it was shown that different types of LSs, both bright and dark, could emerge in Type I SHG doubly resonant cavities \cite{hansson_quadratic_2018}. This study was later extended in \cite{arabi_localized_2020}, where the bifurcation structure of bright quadratic LSs was characterized. The effect that phase mismatch may have on the dynamics of single-peak bright LSs was also analyzed \cite{villois_soliton_2019,erkintalo_dynamics_2019}. In dispersive cavities, LSs are one-dimensional objects that appear along the propagation direction in the cavity.

In this work we analyze the bifurcation structure and stability of quadratic dark LSs arising in doubly resonant cavities enhanced SHG of Type I. Similarly than in Kerr cavities \cite{parra-rivas_dark_2016}, these states emerge due to the locking of plane fronts or domain walls connecting two uniform states of the system in the so called {\it uniform bistable} region. The LSs formed this way undergo a particular bifurcation structure known as {\it collapsed homoclinic snaking} \cite{knobloch_homoclinic_2005,yochelis_reciprocal_2006} which here is analyzed in detail. We show that, although the collapsed snaking is preserved in the parameter space, the stability of the LSs are modified due to the presence of oscillatory instabilities. 

The paper is organized as follows. In Sec.~\ref{sec:1} we present the mean-field model describing our system. In Sec.~\ref{sec:2} we introduce the stationary problem, the spatial dynamical systems associated with it, and describe some of the possible time-independent solutions of the system. After that, Sec.~\ref{sec:3} focuses on the uniform steady states and on their linear stability. Later, Sec.~\ref{sec:4} is devoted to the bifurcation structure and stability of dark LSs. The persistence of these states as varying the control parameters is tackled in Sec.~\ref{sec:5}, where the emergence of breathing behavior is also discussed. Finally, Sec.~\ref{sec:6} ends with a short discussion of the main results and the conclusions. 
\section{The mean-field model}\label{sec:1} 
\begin{figure}[t]
 	\centering
 	\includegraphics[scale=1]{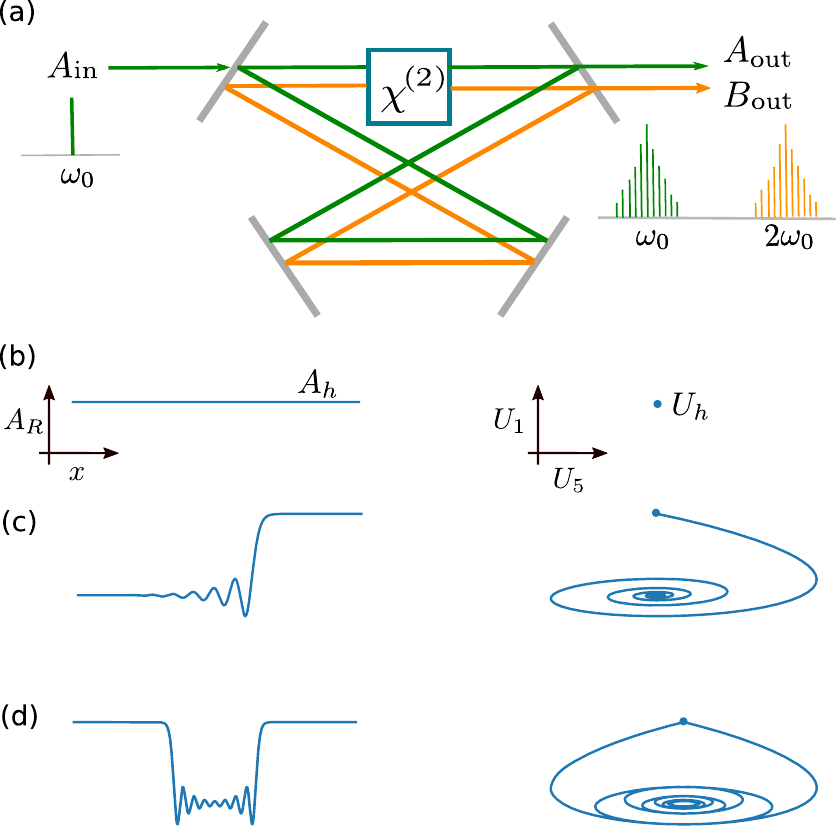}
 	\caption{(a) Schematic view of a cavity enhanced second harmonic generation. The cavity contains a $\chi^{(2)}$-medium and is driven by a continuous-wave field $A_{in}$ at frequency $\omega_0$. The fundamental and generated second-harmonic fields at $2\omega_0$, namely $A$ and $B$, resonate together within the cavity.
 	In panels (b), (c) and (d) we plot the correspondence between the uniform steady states, the plane fronts and a LS states on the left with the fixed point, heteroclinic orbit and homoclinic orbit on the right, respectively.}
 	\label{fig1}
 \end{figure}

Figure~\ref{fig1} shows a schematic example of a doubly resonant dispersive cavity filled with a quadratic nonlinear material ($\chi^{(2)}$). The cavity is driven by a continuous-wave field ($A_{in}$) at frequency $\omega_0$. Inside the cavity, the nonlinear interaction between the electric field and the $\chi^{(2)}$ medium leads, through a process of frequency-up conversion, to the second-harmonic  field (SHF) centered at $2\omega_0$. Then, both the fundamental field (FF) and the SHF resonate in the cavity, and thus we call it doubly-resonant configuration.

In the mean-field approximation (i.e., for high-finesse cavities) the slowly varying envelopes of these fields are described by the nonlinear partial differential equations 
\begin{subequations}\label{MF}
	\begin{equation}\label{MF1}
	\partial_t A=-(1+i\Delta_1)A-i\eta_1\partial_{x}^2A+i B\bar{A}+S
	\end{equation}
	\begin{equation}\label{MF2}
	\partial_t B=-(\alpha+i\Delta_2)B-\left(d\partial_{x}+i\eta_2\partial_{x}^2\right)B+i A^2,
	\end{equation}
\end{subequations}
where $A$ and $B$ are the slowly varying normalized envelopes of FF and SHF respectively \cite{leo_frequency-comb_2016,hansson_singly_2017}. $\bar{A}$ denotes the complex conjugate of $A$.
$t$ corresponds to the normalized slow time describing the evolution of fields after every round-trip, and $x$ is the normalized fast time (in dispersive systems) or space (in diffractive systems). The parameters $\Delta_{1,2}$ are the normalized cavity phase detunings, with $\Delta_2=2\Delta_1+\varrho$ and $\varrho$ the normalized wave-vector mismatch between the fields $A$ and $B$ over one roundtrip; $\alpha$ is the ratio of the round-trip losses $\alpha_{1,2}$ associated with the propagation of FF and SHF; $\eta_{1,2}$ are the normalized group velocity dispersion (GVD) parameters of $A$ and $B$; $d$ is the normalized rate of group velocity mismatch or walk-off between both fields; and $S$ is the normalized driving field amplitude or pump at frequency $\omega_0$. Here $\eta_1=+1$($-1$) denotes normal (anomalous) GVD, and $\eta_2$ can take any positive or negative value. 

Our physical system requires the periodic boundary conditions 
\begin{align*}
A(x+l,t)=A(x,t), && \partial_x A(x+l,t)=\partial_x A(x,t), \\ B(x+l,t)=B(x,t), && \partial_x B(x+l,t)=\partial_x B(x,t),
\end{align*}
with $l$ being the length of the normalized domain (here we fix $l=80$). To perform the analytical computations, however, we consider an infinite domain.  In the following we examine the case with vanishing walk-off ($d=0$). 


\section{The stationary problem}\label{sec:2}
Time-independent states ($\partial_tA=\partial_t B=0$) of the mean-field model (\ref{MF}) satisfy the set of second-order complex differential equations
\begin{subequations}\label{sta_SHG}
	\begin{equation}\label{MF1_sta}
	i\eta_1\partial_{x}^2A=-(1+i\Delta_1)A+i B\bar{A}+S
	\end{equation}
	\begin{equation}\label{MF2_sta}
	i\eta_2\partial_{x}^2B=-(\alpha+i\Delta_2)B+i A^2.
	\end{equation}
\end{subequations}
These states include uniform or homogeneous steady states, spatially extended periodic patterns, stationary plane fronts and LSs. 


To perform a bifurcation analysis of these states we recast Eqs.~(\ref{sta_SHG}) as a finite dimensional dynamical system where  $x$ plays the role of the evolutionary variable \cite{champneys_homoclinic_1998,haragus_local_2011}. 
Taking $A=A_R+iA_I$ and $B=B_R+iB_I$, Eqs.~(\ref{sta_SHG}) are equivalent to the 8D dynamical system
\begin{equation}\label{SD_compact}
\frac{dU}{dx}=f\left(U;\Delta_i,\eta_i,S\right).
\end{equation}
with the new variable
\begin{equation*}
U\equiv\left(\begin{array}{l}
U_1\\U_2\\U_3\\U_4\\U_5\\U_6\\U_7\\U_8
\end{array}\right)=
\left(\begin{array}{l}
A_R\\
A_I\\
B_R\\
B_I\\
\partial_xA_R\\
\partial_xA_I\\
\partial_xB_R\\
\partial_xB_I.
\end{array}\right),
\end{equation*}
and the vectorial field $f$ defined as 
\begin{equation*}
f\equiv\left(\begin{array}{l}
f_1\\f_2\\f_3\\f_4\\f_5\\f_6\\f_7\\f_8
\end{array}\right)=
\left(\begin{array}{l}
U_5\\
U_6\\
U_7\\
U_8\\
\eta_1^{-1}[-(\Delta_1U_1+U_2)+U_3U_1+U_4U_2]\\
\eta_1^{-1}[U_1-\Delta_1U_2-U_3U_2+U_4U_1-S]\\
\eta_2^{-1}[-\alpha U_4-\Delta_2U_3+U_1^2-U_2^2]\\
\eta_2^{-1}[\alpha U_3-\Delta_2U_4+2U_1U_2]
\end{array}\right).
\end{equation*}
This approach, known as spatial dynamics, is very useful when dealing with coherent states such as LSs, as it permits applying well known results of dynamical systems and bifurcation theory to study (time-independent) localized solutions. 

In this framework, a correspondence between solutions of Eq.~(\ref{sta_SHG}) and Eq.~(\ref{SD_compact}) can be established such that a uniform state of Eq.~(\ref{sta_SHG}) corresponds to a fixed point of Eq.~(\ref{SD_compact}) [see Fig.~\ref{fig1}(b)]; a plane front wave connecting two uniform states consists in a heteroclinic orbit [see Fig.~\ref{fig1}(c)], and a LS to a homoclinic orbit [see Fig.~\ref{fig1}(d)].

This system is spatially reversible, that is, invariant under the transformation $x\rightarrow -x$. This reflection symmetry allows us to compute such states as solutions of Eq.~(\ref{SD_compact}) considering just half of the domain $[0,l/2]$ and the Neumann boundary conditions
$U_5(0)=U_6(0)=U_7(0)=U_8(0)=0,$ and $U_5(l/2)=U_6(l/2)=U_7(l/2)=U_8(l/2)=0$.

Afterwards, we compute the linear temporal stability of these states solving the eigenvalue equation 
\begin{equation}\label{linsta}
\mathcal{L}\psi=\sigma\psi,
\end{equation} 
where $\mathcal{L}$ is the linear operator associated with the right hand side of Eq.~(\ref{MF}), once evaluated at a given steady state, and $\sigma$ is the eigenvalue associated with the eigenfunction $\psi$, both depending on the control parameters of the system. 
\section{Uniform steady state and linear stability}\label{sec:3}
The first step in our analysis is to characterize the uniform or homogeneous steady state of the system $(A_h,B_h)$, i.e., the continuous-wave of the system. The modification of this state and its stability in the parameter space allow us to predict the type of LSs which may emerge in these type of devices.

\begin{figure}[!t]
 	\centering
 	\includegraphics[scale=1]{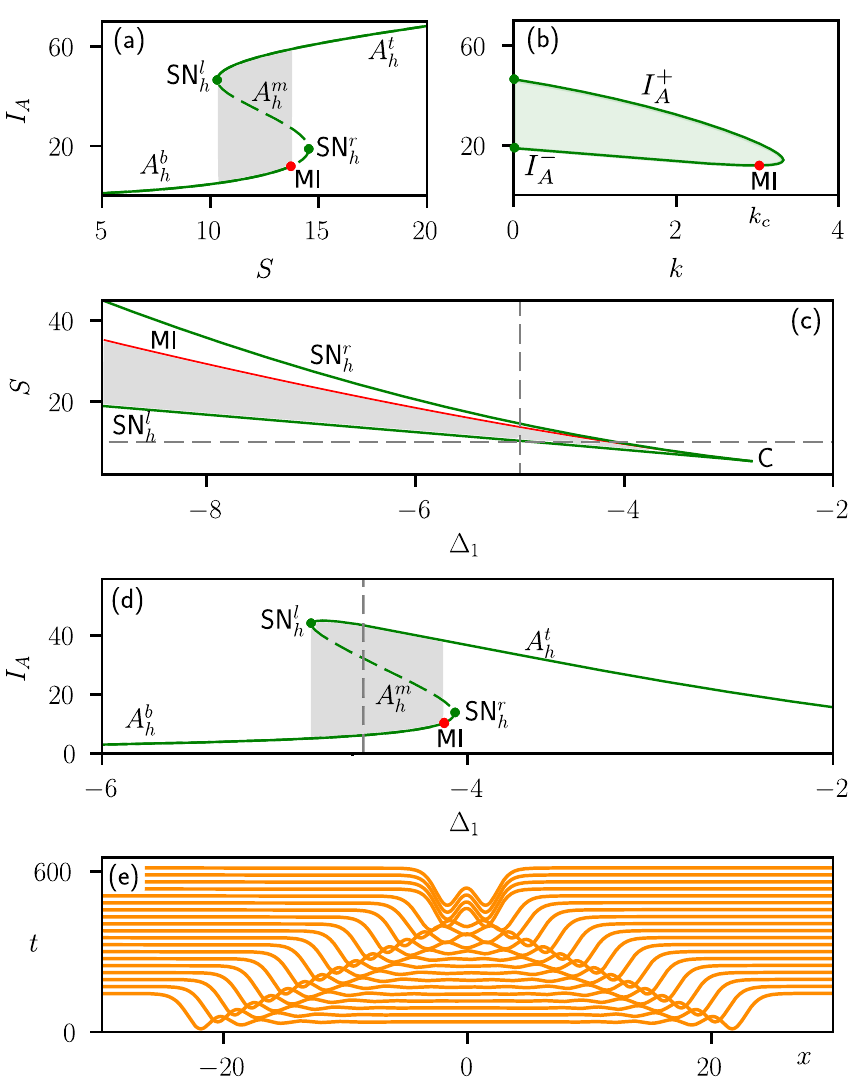}
 	\caption{(a) Uniform state for $(\alpha,\Delta_1,\varrho)=(1,-5,0)$. Stable (unstable) branches are plotted with solid (dashed) lines. (b) Marginal stability curve computed for $(\eta_1,\eta_2)=(-1,1/2)$. (c) Phase diagram in the $(\Delta_1,S)$-parameter plane showing SN$_h^{l,r}$ and the MI line. (d) Uniform state as a function of $\Delta_1$ for $S=10$. In panels (a), (c) and (d) the uniform bistable region is shown in gray. Panel (e) shows the interaction and locking of two plane fronts for $(\Delta_1,S)=(-4.57,10)$ [see vertical dashed line in panel (d)]. }
 	\label{fig2}
 \end{figure}

These states satisfy $\partial_xA_h=\partial_xB_h=0$, and are solutions of the equations
 \begin{subequations}
 	\begin{equation}\label{hom_A}
 	-(1+i\Delta_1)A_h+(\alpha+i\Delta_2)^{-1}|A_h|^2A_h+S=0 
 	\end{equation}
 	\begin{equation}
 	B_h=-i(\alpha+i\Delta_2)^{-1}A_h^2. 
 	\end{equation}
 \end{subequations}
$B_h$ is slaved to $A_h$, and thus, the uniform steady state is completely defined through only the FF $A_h$. Equation~(\ref{hom_A}) can be rewritten in terms of the intensity $I_A=|A_h|^2$ as  
\begin{equation}\label{Shom}
S^2=\frac{(1+\Delta_1^2)(1+\Delta_2^2)I_A+2\alpha(1-\Delta_1\Delta_2)I_A^2+I_A^3}{\alpha^2(1+\Delta_2^2)}.
\end{equation}
According to the previous equation, our system may exhibit a hysteresis cycle like the one shown in Fig.~\ref{fig2}(a) for $(\alpha,\Delta_1,\varrho)=(1,-5,0)$. Here, the system has three uniform solutions labeled $A_h^b$, $A_h^m$ and $A_h^t$, which are separated by the two folds occurring at
 \begin{equation}\label{eq_folds}
 I_A^{t,b}=\frac{\alpha}{3}\left(2(\Delta_1\Delta_2-1)\pm\sqrt{g}\right),
 \end{equation}
 where $g\equiv(\Delta_2^2-3)\Delta_1^2-3\Delta_2^2-8\Delta_1\Delta_2+1$.
  \begin{figure*}[!t]
  	\centering
  	\includegraphics[scale=1.05]{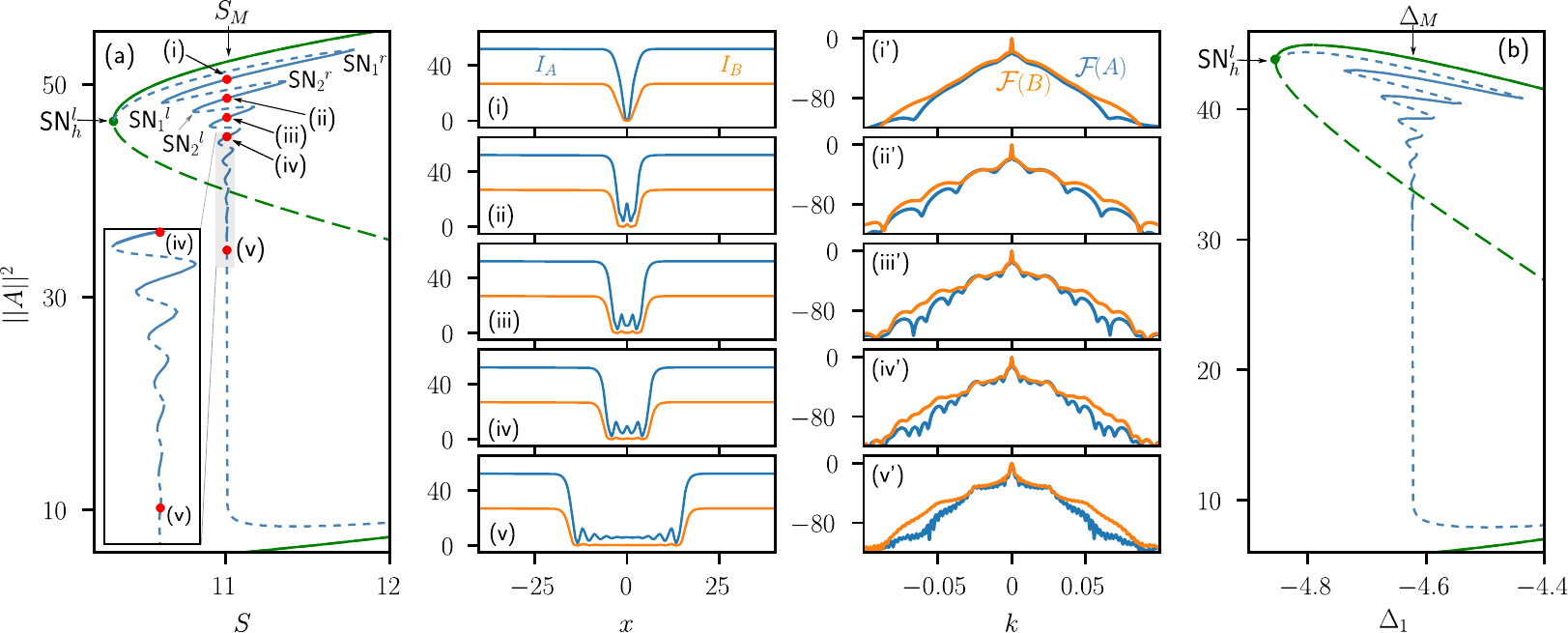}	
  	\caption{
  		(a) Collapsed snaking bifurcation diagram as a function of $S$ for  $\Delta_1=-5$. Stable (unstable) branches are plotted with solid (dashed) lines. A close-up view of this structure is shown in the inset. This bifurcation diagram corresponds to the vertical dashed line shown in Fig.~\ref{fig4}. Labels (i)-(v) correspond to the stable LSs shown on the right. Panels (i')-(v') show the Fourier transform ($\mathcal{F}$) in logarithmic scale of the LSs shown respectively in panels (i)-(v). (b) Collapsed snaking diagram as a function of $\Delta_1$ for $S=10$. It corresponds to the horizontal dashed line shown in Fig.~\ref{fig4}.   Here,  $(\alpha,\varrho,\eta_1,\eta_2)=(1,0,-1,1/2)$.  }
  	\label{fig3}
  \end{figure*}  
The next step in our analysis consists in determining the linear stability of $(A_h,B_h)$ against small modulated plane-wave perturbations $\psi_k(x,t)=\xi e^{\sigma t-ikx}+c.c.$, where  $\sigma$ and $k$ are the growth rate and the wavenumber of the perturbation, and $\xi$ is a small amplitude field. We insert weakly modulated states
of the form  $A(x,t)=A_h+\psi_k(x,t)$
(equivalently for $B$) into Eq.~(\ref{MF}) and linearize the system around $(A_h,B_h)$. This yields the eigenvalue problem (\ref{linsta}) from where one can obtain $\sigma$ analytically. 

In the absence of walk-off, this calculation leads to the dispersion relation
\cite{trillo_pulse-train_1996,hansson_quadratic_2018}:  
\begin{equation}
\sigma=-1\pm\sqrt{-(f_1+f_2)/2\pm\sqrt{p+(f_1-f_2)^2/4}},
\end{equation}
where
\begin{align*}
f_1=\tilde{\Delta}_1^2+2I_A-I_B, && f_2=\tilde{\Delta}^2_2+2I_A,
\end{align*}
\begin{align*}
p=2I_A[(\tilde{\Delta}_1+\tilde{\Delta}_2)^2-I_B], &&\tilde{\Delta}_{1,2}\equiv\Delta_{1,2}-\eta_{1,2}k^2, 
\end{align*}
and $I_B\equiv|B_h|^2$. The uniform state $(A_h,B_h)$ is stable (unstable) whenever Re$[\sigma]<0$ (Re$[\sigma]>0$). The transition between these two situations occurs at the Turing, or modulational instability (MI), which takes place when the two conditions $\sigma(k)|_{k_c}=0$ and $d\sigma/dk|_{k_c}=0$ are satisfied. 

The stability range of the uniform state is determined by the condition $\sigma=0$, which leads to the equation
\begin{equation}
 c_2 I^2_A+c_1 I_A+c_0=0,
\end{equation} 
 where
\begin{align*}
 	c_0=\tilde{\Delta}_1^2+\tilde{\Delta}_2^2+\tilde{\Delta}_2^2\tilde{\Delta}_1^2+1,&& 	c_1=4(1-\tilde{\Delta}_1\tilde{\Delta}_2),
\end{align*} 
and 
 	\begin{equation*}
c_2=4-\frac{1-\tilde{\Delta}_2^2}{\alpha^2+\Delta_2^2}
 	\end{equation*}
This equation support to solution curves $I_A^\pm$ which together define the marginal instability curve (MIC). An example of the MIC and the stability region associated with the uniform states shown in Fig.~\ref{fig2}(a) are plotted in Fig.~\ref{fig2}(b) for $(\eta_1,\eta_2)=(-1,1/2)$. For a fixed value of $k=k_*$, the uniform state whose intensity $I_A$ lies inside the MIC (i.e., $I_A^-<I_A<I_A^+$) is temporally unstable, and stable otherwise. Thus, all the area in-between $I_A^\pm$ (the {\it unstable tongue}) defines the unstable uniform states. In correspondence, Fig.~\ref{fig2}(a) shows the stable (unstable) states using solid (dashed) lines. The minimum of this unstable tongue occurs at $(I_A,k)=(I_c,k_c)$, and signals the position of the MI [see red dot in Fig.~\ref{fig2}(a)]. For $k=0$, two saddle-node (SN) bifurcations [see green dots in Fig.~\ref{fig2}(b)] emerge, and correspond to the folds points defined by Eq.~(\ref{eq_folds}). In what follows we denote those bifurcations SN$_h^{l,r}$. 

The bifurcation diagram shown in Fig.~\ref{fig2}(a) corresponds to a slice at constant $\Delta_1$ of the $(\Delta_1,S)-$phase diagram in Fig.~\ref{fig2}(c) (see dashed vertical line). The saddle-nodes in Fig.~\ref{fig2}(a) separate with increasing $|\Delta_1|$, while they come closer with decreasing it, colliding eventually in a cusp $C$ when $g=0$. Below this point, the uniform state is monotonous with $S$. The MI corresponds to the red line plotted in Fig.~\ref{fig2}(c).
In the region between MI and SN$_h^l$, $A_h^b$ and $A_h^t$ coexist and are stable [see gray shadowed area in Figs.~\ref{fig1}(a),(c)]. We refer to this region as the {\it uniform bistability} region.

For $\varrho=0$, the solutions of Eq.~(\ref{Shom}) can be also plotted as a function of $\Delta_1$ using the relation 
\begin{equation}
\Delta_1=\pm\sqrt{\frac{b^2\pm\sqrt{b^2-9I_Ac}}{8I_A}},
\end{equation}
with 
\begin{align*}
b=5I_A-4\alpha I_A^2-4\alpha^2S^2, && c=I_A-\alpha^2S+2\alpha I_A^2+I_A^3.
\end{align*}
An example of the uniform state as a function of $\Delta_1$ is shown for $S=10$ in Fig.~\ref{fig2}(d). This curve is the nonlinear resonance of the cavity, and corresponds to the horizontal dashed line plotted in Fig.~\ref{fig2}(c).

\section{Bifurcation structure and stability of dark states}\label{sec:4}
Within the uniform bistability region shown in Fig.~\ref{fig2}, plane fronts like those shown in Fig.~\ref{fig1}(c) can emerge, connecting the uniform states $A_h^b$ with $A_h^t$. We distinguish two different polarities, depending on whether the connection is from $A_h^b$ to $A_h^t$ (up) or vice-versa (down). These fronts move at a constant speed which depends on the control parameters of the system. As a result, two fronts of different polarity will come closer of move apart, depending on their velocities.  

The front speed cancels out in a single set of parameters known as the Maxwell point of the system \cite{pomeau_front_1986}.
Around this point, fronts can lock one another through the overlapping of their oscillatory tails, leading to the formation of LSs of different extension \cite{,coullet_localized_2002,coullet_nature_1987}.  An example of the evolution of the fronts and locking process is shown in Fig.~\ref{fig2}(e) for $A_R$ and  $(\Delta_1,S)=(-4.57,10)$.

\begin{figure}[!t]
	\centering
	\includegraphics[scale=1]{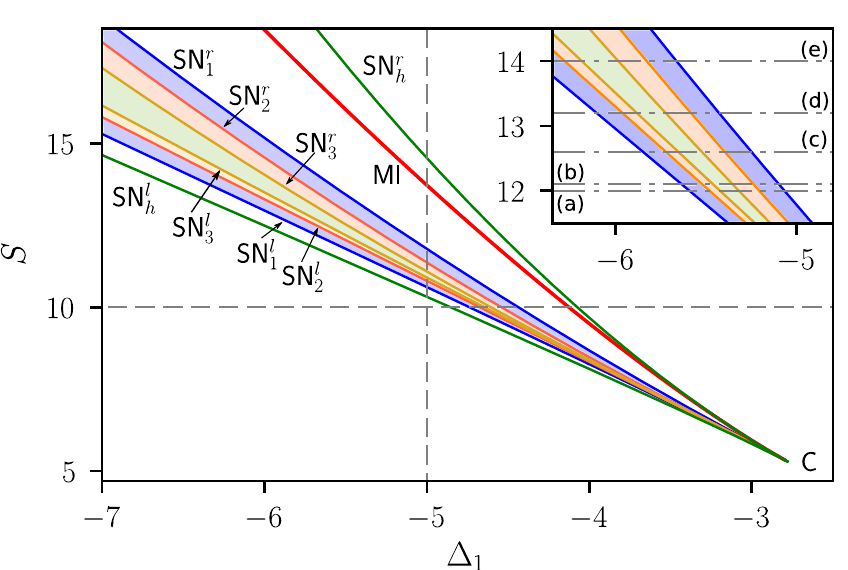}
	\caption{Phase diagram in the $(\Delta_1,S)$-parameter space for $(\alpha,\varrho,\eta_1,\eta_2)=(1,0,-1,1/2,0)$, showing the main bifurcation lines of the system: SN$_h^{l,r}$, SN$_{1,2,3}^{l,r}$, and MI. The vertical and horizontal dashed lines correspond to the diagrams shown in Fig.~\ref{fig3}(a) and \ref{fig3}(b) respectively. The point-dashed lines in the insets correspond to the bifurcation diagrams shown in Fig.~\ref{fig5} }
	\label{fig4}
\end{figure}

\begin{figure*}[!t]
\centering
\includegraphics[scale=1.05]{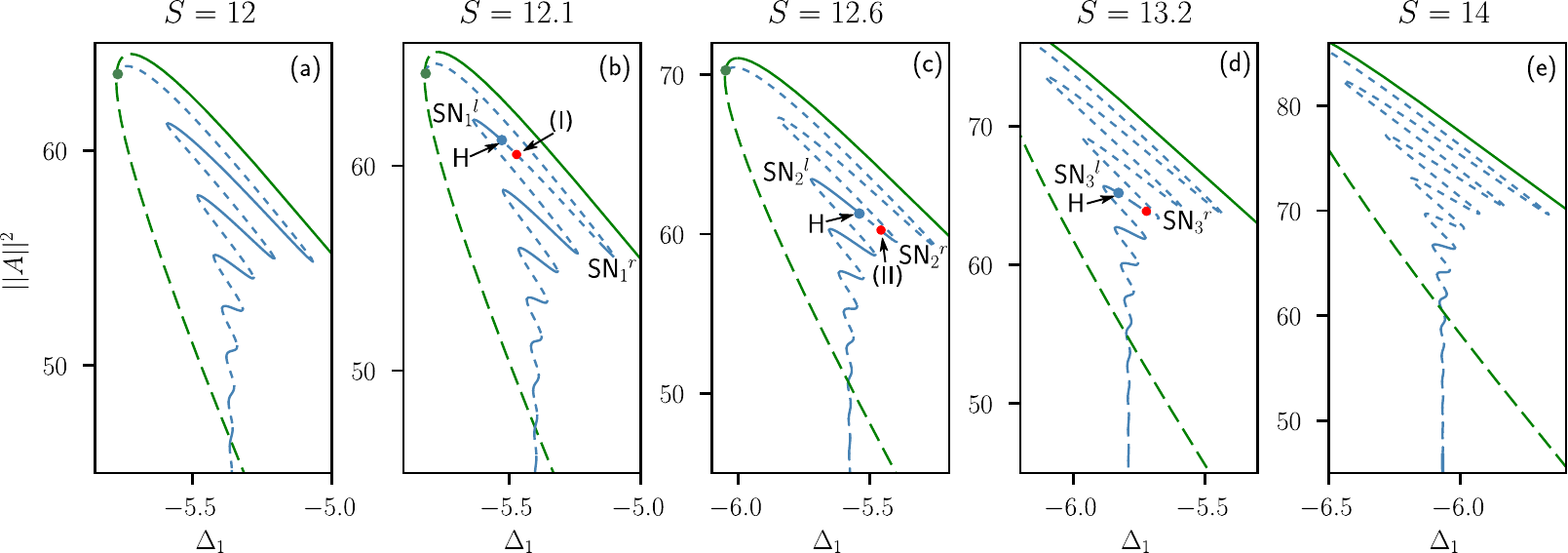}
\caption{Destabilization of LSs through Hopf instabilities. The bifurcation diagrams show the modification of the collapsed snaking with increasing $S$. Stable (unstable) state branches are plotted using solid (dashed) lines. In (a) the stability is preserved. In (b) the branch of the single-dip dark state is partially destabilized through a Hopf bifurcation (H). An example of a breather for $\Delta_1=-5.47$ [see red dot with label (I)] is shown in Fig.~\ref{fig6}(I).
Panel (c) shows the whole branch of the single-dip states completely destabilized, and the stable branch associated with the two-dip dark state is partially Hopf unstable. An example of an oscillatory state for $\Delta_1=-5.52$ [see red dot labeled (II)] is shown in Fig.~\ref{fig6}(II). In panel (d) the Hopf bifurcation has moved down, and the stable branch associated with the three-dip dark LS is partially unstable. In panel (e) branches of wider LSs are now oscillatory unstable. 
}
\label{fig5}
\end{figure*}
\begin{figure*}[!t]
	\centering
	\includegraphics[scale=1]{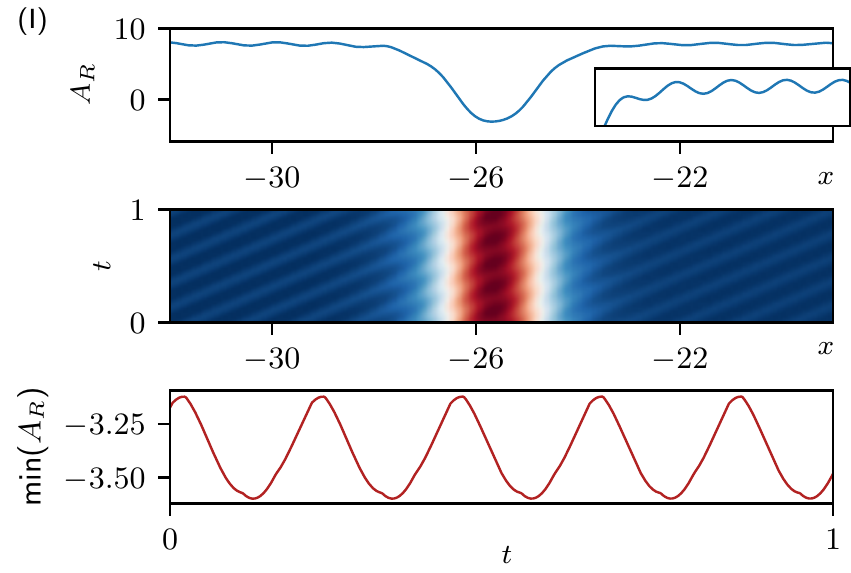}
		\includegraphics[scale=1]{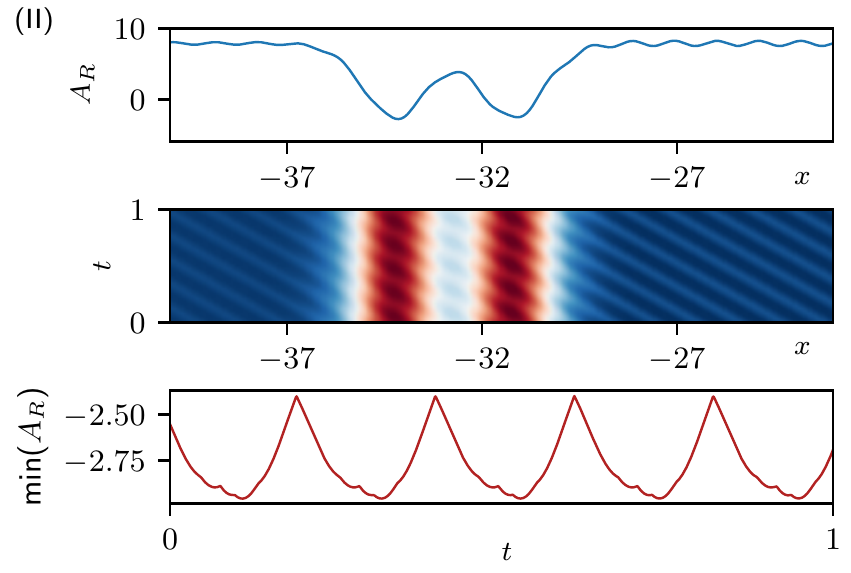}
	\caption{Breathing behavior undergone by dark LSs of different extension. Panel (I) shows the spatio-temporal evolution of $A_R$ associated with a single-dip dark breather [see top and middle panels], and the time evolution of its minimum [bottom panel] for $(\Delta_1,S)=(-5.47,12.1)$. In panel (II) same than in (I) for a two-dip dark breather for $(\Delta_1,S)=(-5.52,12.6)$.}
	\label{fig6}
\end{figure*}
To fully understand the emergence of these states, it is necessary to compute their bifurcation structure and determine their stability. To do so we apply a numerical path-continuation procedure based on a Newton-Raphson solver \cite{doedel_numerical_1991-1,doedel_numerical_1991,allgower_numerical_1990}, and  use the free distribution software package AUTO-07p \cite{Doedel2009}. In what follows we consider 
$\alpha=1$, vanishing phase-mismatch ($\varrho=0$), and we fix $\eta_1=-1$, and $\eta_2=1/2.$ 

 The outcome of these computations yields the bifurcation diagram shown in Fig.~\ref{fig3}(a), where the energy of $A$, i.e. the $L_2$-norm 
$$||A||^2=\frac{1}{l}\int_{-l/2}^{l/2}|A(x)|^2dx$$
is plotted as a function of the amplitude of the driving field, $S$.

The bifurcation curve undergoes a damped oscillation in the parameter $S$ which collapses, with decreasing $||A||^2$, asymptotically to the Maxwell point of the system at $S=S_M$. 
Along this bifurcation curve, known as {\it collapsed homoclinic snaking} \cite{knobloch_homoclinic_2005,yochelis_reciprocal_2006,parra-rivas_dark_2016}, the LSs modify their morphology and stability. A few examples of stable LSs are shown in Fig.~\ref{fig3}, along with their Fourier transforms. The stability of the different state branches are marked using solid (dashed) lines for stable (unstable) solutions, and has been obtained by solving Eq.~(\ref{linsta}) numerically.

For high values of $||A||^2$, the bifurcation curve arise from SN$_h^l$. Near SN$_h^l$, the LS consists in a small amplitude dip. In the weakly nonlinear regime, i.e., very close to SN$_h^l$, these states can be captured by the asymptotic solution $A(x)-A_h\sim a{\rm sech}^2(bx)$, where the coefficients $a,b$ depend on the parameters of the system \cite{burke_classification_2008,parra-rivas_dark_2016}.

LSs emerge unstably from SN$_h^l$, and increase their amplitude as
we follow the diagram towards 
the first fold on the right. This fold corresponds to the saddle-node bifurcation SN$_1^r$. At this bifurcation, the single-dip LS stabilizes preserving such stability until reaching SN$_1^l$. An example of this profile is shown in Fig.~\ref{fig3}(i).

Once SN$_1^l$ is passed and $S$ is increased, the single-dip dark state develops a small bump at $x=0$, and at SN$_2^r$ it becomes stable, looking like the dark LS shown in Fig.~\ref{fig3}(ii). Decreasing $||A||^2$, the bump nucleation process repeats and the LSs become wider and wider as the bifurcation curve collapses to $S_M$ [see for example profiles (iii)-(iv) in Fig.~\ref{fig3}]. At this stage, the LS is like the one shown in Fig.~\ref{fig3}(v) where two well separated plane fronts can be distinguished. Decreasing the energy below $||A||^2\approx 10$, the bifurcation curve separates from $S_M$, and eventually connects back to $A_h^b$ at the MI (not shown here).

Performing a two-parameter continuation, we are able to compute the modification of the saddle-node bifurcations shown in Fig.~\ref{fig3}(a) with varying $\Delta_1$. As a result, we obtain the phase diagram in the $(\Delta_1,S)$-parameter plane shown in Fig.~\ref{fig4}. In this diagram, the main bifurcation lines of the system, namely SN$_h^{l,r}$, MI, and SN$_{1,2,3}^{l,r}$ are depicted. Decreasing $|\Delta_1|$, the pair of saddle-nodes SN$_i^{l,r}$ approach one another and eventually collide in a sequence of cusp bifurcations $C_i$, where wider LSs disappears sequentially. Eventually, $C_1$ occurs, and LSs completely disappear. Increasing $|\Delta_1|$ however, the extension of the bistability region widens and so does the region of existence of different types of LSs. The bifurcation diagram shown in Fig.~\ref{fig3}(a) corresponds to a slice of the phase diagram in Fig.~\ref{fig4} for constant $\Delta_1=-5$ (see vertical dashed line). 

One could also understand this diagram in a different way by taking slices of constant $S$ and letting $\Delta_1$ vary. This perspective is easier to understand for experimentalists who scan the cavity in $\Delta_1$ while fixing the intensity of the driving source. Thus, 
for completeness, we also show the collapsed snaking structure that one obtains when fixing $S$ and modifying $\Delta_1$. Figure~\ref{fig3}(b) shows such structure for $S=10$ corresponding to the horizontal dashed line plotted in Fig.~\ref{fig4}. For this value, the stability of the different branches is similar to the one shown in Fig.~\ref{fig3}(a).

\section{Persistence of the collapsed snaking and breathing behavior}\label{sec:5}
The collapsed snaking structure persists within the whole uniform bistability region, as shown in the $(\Delta_1,S)$-phase diagram of  Fig.~\ref{fig4}. However, the stability of their associated LSs modify with increasing $S$, through the appearance of oscillatory or Hopf bifurcations. The stability variation of the collapsed snaking is shown in Fig.~\ref{fig5} for five values of $S$, which correspond to the horizontal dashed lines plotted in the inset of Fig.~\ref{fig4}.

For $S=12$ [see Fig.~\ref{fig5}(a)], the stability of the different branches is the same as the one shown in Fig.~\ref{fig3}(b) for $S=10$. Increasing $S$ a bit further, however, a Hopf (H) bifurcation pops up from SN$_1^r$, and moves towards SN$_1^l$ as $S$ grows. One example of this configuration is plotted in Fig.~\ref{fig5}(b) for $S=12.1$. Within the region between H and SN$_1^r$ the static single-dip LS is unstable, and breathers like the one shown in Fig.~\ref{fig6}(I) arise. The particularity of these states is that, as soon as they start breathing, they develop a modulated background of small spatial period which drift to the right at a constant speed [see Fig.~\ref{fig6}(I)]. Furthermore, the oscillations of the LS are not left/right symmetric, but there is a displacement of the minimum from left to right, which can be appreciated in Fig.~\ref{fig6}(I). The LS as a whole also drifts to the right albeit at a much slower constant speed than the modulated background (not shown here). These states breathe with a single period, as shown by the temporal variation of its minimum [i.e., min($A_R$)].
Increasing $S$ a bit further, H  eventually collides with SN$_1^l$ in a codimension-two bifurcation, destabilizing completely the branch associated with the single-dip LSs.

Figure~\ref{fig5}(c) shows the bifurcation diagram for $S=12.6$. In this case, the stable branch associated with the two-dip LS becomes partially unstable through another Hopf bifurcation.  Similarly, this point emerges from SN$_2^r$, moves towards SN$_2^l$ with increasing $S$, and eventually dies out at the fold. The breather here is like the one depicted in Fig.~\ref{fig6}(II), where now the two dips and central bump oscillate asymmetrically with a single period and drift at constant speed to the left, together with the much faster background periodic modulation.

Progressing up in the phase diagram of Fig.~\ref{fig4}, the H moves down along the collapsed snaking curve. The diagram plotted in  Fig.~\ref{fig5}(d) shows the situation for $S=13.2$ where the three-dip dark state branch is partially unstable. While increasing $S$, the Hopf bifurcation moves down in the diagram destabilizing branches of wider LSs [see for example Fig.~\ref{fig5}(e) for $S=14$]. Eventually, for even larger values of the pump, all the collapsed snaking become unstable, and only breathers exist.   

The dynamics of breathers in static or dynamic backgrounds has been studied by different authors in conservative systems \cite{Xue_2020,pelinovsky_localized_2020,zhou2021breathers}. The emergence of traveling pulses on a periodically modulated background has been also studied in dissipative systems \cite{PhysRevE.91.050901}, although their origin is very different to the one reported here and do not breathe. Therefore, the emergence of the breather states shown here are, as far as we known, largely unexplored. One plausible explanation may be related with the asymmetry of the Hopf mode associated with the breather. However, a complete understanding of this phenomenon requires further study.

\section{Discussion and conclusions}
\label{sec:6}
In this paper we have presented a detailed bifurcation analysis of dark LSs arising in doubly resonant second-harmonic dispersive cavities. In the high-finesse limit, these cavities are described by a mean-field model for FF and SHF (Sec.~\ref{sec:1}). We have mainly focused on time-independent states which are described by the stationary version of the mean-field model or its equivalent spatial dynamical system (Sec.~\ref{sec:2}). The simplest time-independent state is the uniform steady state (i.e., the continuous-wave of the system). The linear stability of these states gives plenty of information which can anticipate the potential types of LSs and their bifurcation structure. Indeed, such analysis reveals that, for the parameter region chosen in this work, the system presents bistability between two coexisting uniform states (see Sec.~\ref{sec:3}). In this context, plane front waves may arise, interact and eventually lock, forming dark LSs of different extensions. These states undergo collapsed homoclinic snaking: a damped oscillating bifurcation curve in $S$ (or $\Delta_1$)  around the Maxwell point which asymptotically approaches it as the width of the LSs increases, i.e. with decreasing the energy.
Similar bifurcation structures also arise in Kerr cavites, optical parametric oscillators, and cavity enhanced SHG in a different regime  \cite{parra-rivas_dark_2016,parra-rivas_localized_2019,arabi_localized_2020}. In particular, the scenario described here is qualitatively identical to the one reported in the context of dark LSs in Kerr cavities \cite{parra-rivas_dark_2016}.

The collapsed snaking persists in the $(\Delta_1,S)$-parameter regime, although the stability of the LSs is modified (Sec.~\ref{sec:4}). Indeed, increasing $S$, a Hopf bifurcation crops up, destabilizing 
LSs sequentially from higher to lower energy. As a result,
complex breather behavior arises where the localized oscillations slowly drift on a periodically modulated background traveling at a much faster speed. Eventually, all the static LSs become unstable to breathing ones.

Another important question that definitely must be tackled in future investigations is the impact of a non-vanishing walk-off ($d\ne0$) on the present states. Preliminary analysis shows that for a weak walk-off the bifurcation structure and stability are preserved despite the asymmetry induced by that term. However, a more detailed study is required. 


\bibliographystyle{ieeetr}
\bibliography{LSs_SHG}

\end{document}